\documentstyle[aps,manuscript,epsfig]{revtex}
\begin{document}
\title{
Magnetic ordering and fluctuation in kagom\'{e}
 lattice antiferromagnets,
Fe and Cr jarosites
}
\author{M. Nishiyama, T. Morimoto and S. Maegawa}
\address{Graduate School of Human and Environmental Studies, Kyoto
University, Kyoto, 606-8501, Japan}
\author{ T. Inami}
\address{Department of Synchrotron Radiation Research, Japan Atomic Energy
Reseach Institute, Mikazuki, Hyogo 679-5148, Japan}
\author{Y. Oka}
\address{Faculty of Integrated Human Studies, Kyoto University, Kyoto,
606-8501, Japan}
%
%
%
\maketitle
\newpage
\begin{abstract}
Jarosite family compounds,
KFe$_3$(OH)$_6$(SO$_4$)$_2$, (abbreviate Fe jarosite),
and KCr$_3$(OH)$_6$(SO$_4$)$_2$, (Cr jarosite),
are typical examples of the Heisenberg antiferromagnet on the kagom\'{e}
lattice and have been investigated by means of magnetization and NMR
experiments.
The susceptibility of Cr jarosite deviates from Curie-Weiss law due to
the short-range spin correlation below about $150\mbox{\,K}$ and shows the
magnetic transition at $4.2\mbox{\,K}$,
while Fe jarosite has the transition at $65 \mbox{\,K}$.
The susceptibility data fit well with the
calculated one on the high temperature expansion for the Heisenberg
antiferromagnet on the kagom\'{e} lattice .
The values of exchange interaction of Cr jarosite and Fe jarosite are
derived to be $J_{\rm{Cr}}\!=\!4.9 \,{\rm K}$ and $J_{\rm{Fe}}\!=\!23
\,{\rm K}$, respectively.
The $^1$H-NMR spectra of Fe jarosite suggest that
the ordered spin structure is the ${\bf q}\!=\!0$ type with positive
chirality of the 120$^{\circ}$ configuration.
The transition is caused by a weak single-ion type anisotropy.
The spin-lattice relaxation rate, $1/T_1$, of Fe jarosite in the ordered
phase decreases sharply with lowering the temperature and can be well
explained by the two-magnon process of spin wave with the anisotropy.
\end{abstract}
\newpage
\section{INTRODUCTION}
The antiferromagnets on the kagom\'{e} lattice have frustration due to the
competition of the antiferromagnetic interactions between neighboring spins.
The antiferromagnets on the triangular lattice also have been well known as
the frustration systems.
While the triangular lattice has 6 nearest neighbors and the adjacent
triangles on the triangular lattice share one side,
or 2 lattice points,
in common,
the kagom\'{e} lattice has only 4 nearest neighbors and the adjacent
triangles on the kagom\'{e} lattice share only one lattice point in common.
Thus the spins on the kagom\'{e} lattice suffer smaller restriction from
neighboring spins than the spins on the triangular lattice.
The Heisenberg antiferromagnet on the kagom\'{e} lattice exhibits infinite
and continuous degeneracy of the ground state.
Theoretically the two-dimensional isotropic Heisenberg kagom\'{e} lattice
antiferromagnet have no magnetic phase transition at finite temperature.
The thermal or the quantum fluctuation, however, resolves the degeneracy of
the ground state \cite{RefReimers,RefChubukov}.
This effect induces the coplanar spin arrangement and two N\'{e}el states
have been discussed as candidates for the spin structure at zero
temperature.
One is a ${\bf q }\!=\!0$ type and the other is a $\sqrt{3} \! \times \!
\sqrt{3}$ type of the 120$^\circ$ structure.
Theoretical studies suggest that the latter is favored slightly
\cite{RefReimers}.
When a weak Ising-like anisotropy is introduced into the Hisenberg
kagom\'{e} lattice antiferromagnet,
the system has the magnetic phase transition at finite temperature and has
a peculiar spin structure \cite{RefKuroda}.
Small perturbation, anisotropy or distortion may resolve the degeneracy of
frustrated systems and cause the phase transition.

The jarosite family compounds, KCr$_3$(OH)$_6$(SO$_4$)$_2$ and
KFe$_3$(OH)$_6$(SO$_4$)$_2$ are examples of the Heisenberg kagom\'{e}
lattice antiferromagnets \cite{RefTakano,RefTownsend}.
We have investigated these powder samples by the measurements of
magnetization and the $^1$H nuclear magnetic resonance experiments to
clarify the magnetic transition and the spin fluctuation in the Heisenberg
kagom\'{e} lattice antiferromagnet \cite{RefMaegawa}.
The magnetic ions Cr$^{3+}$ and Fe$^{3+}$ have spins 3/2 and 5/2, respectively.
The ions form the kagom\'{e} lattice on the $c$-plane and interact
antiferromagnetically with each other.
The protons observed by NMR locate nearly on the kagom\'{e} planes.
Adjacent kagom\'{e} planes are separated by nonmagnetic ions, S, O, H and K
with the long interaction paths, so that the interplane magnetic
interaction is very weak.
\section{EXPERIMENTAL RESULTS}
\begin{figure}
\vspace{0cm}
\centerline{\epsfig{file=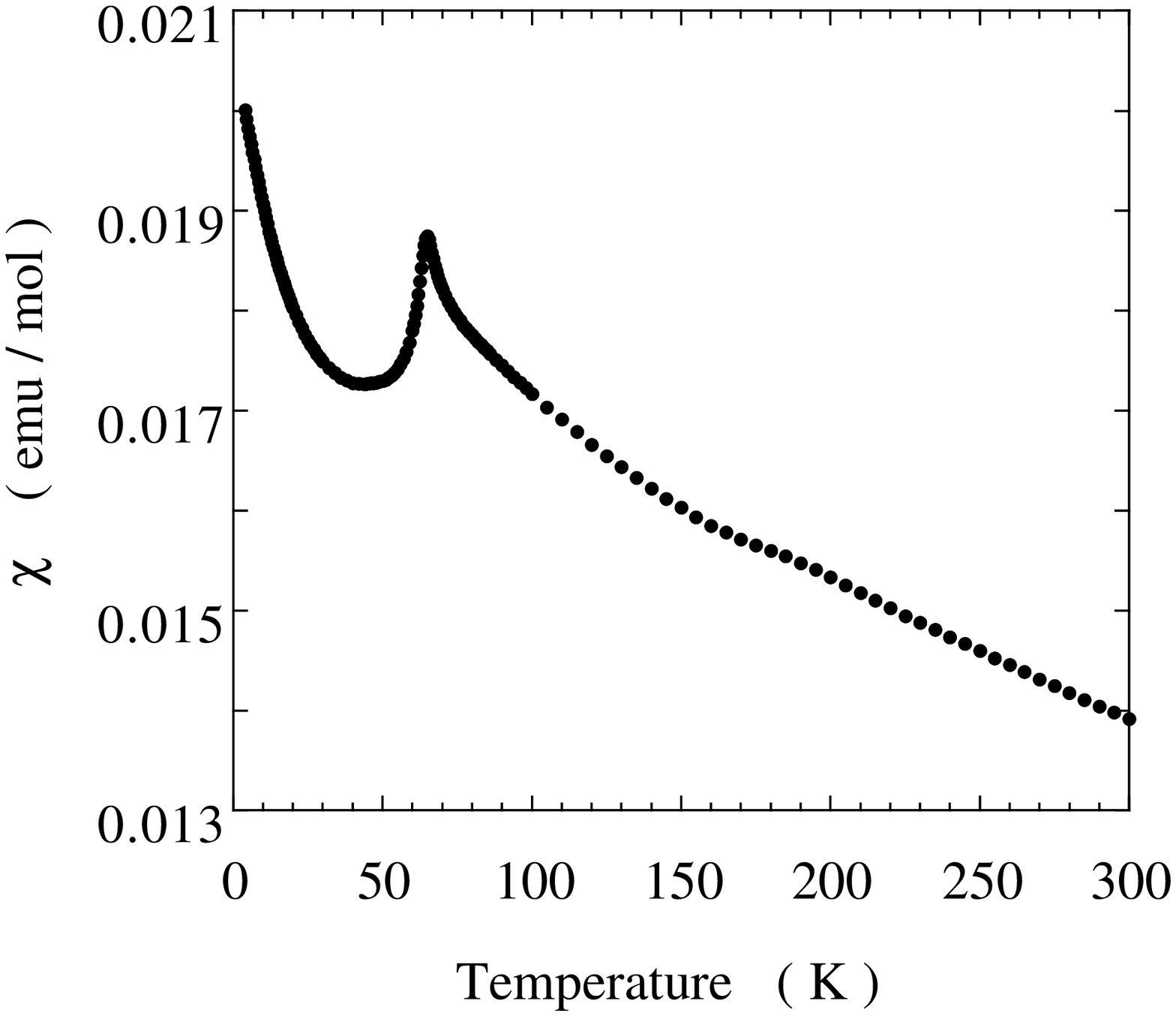,width=16cm}}
\vspace{0.5cm}
\caption{Temperature dependence of the susceptibility for
KFe$_3$(OH)$_6$(SO$_4$)$_2$.}
\label{MNFig1}
\end{figure}

\begin{figure}
\vspace{0cm}
\centerline{\epsfig{file=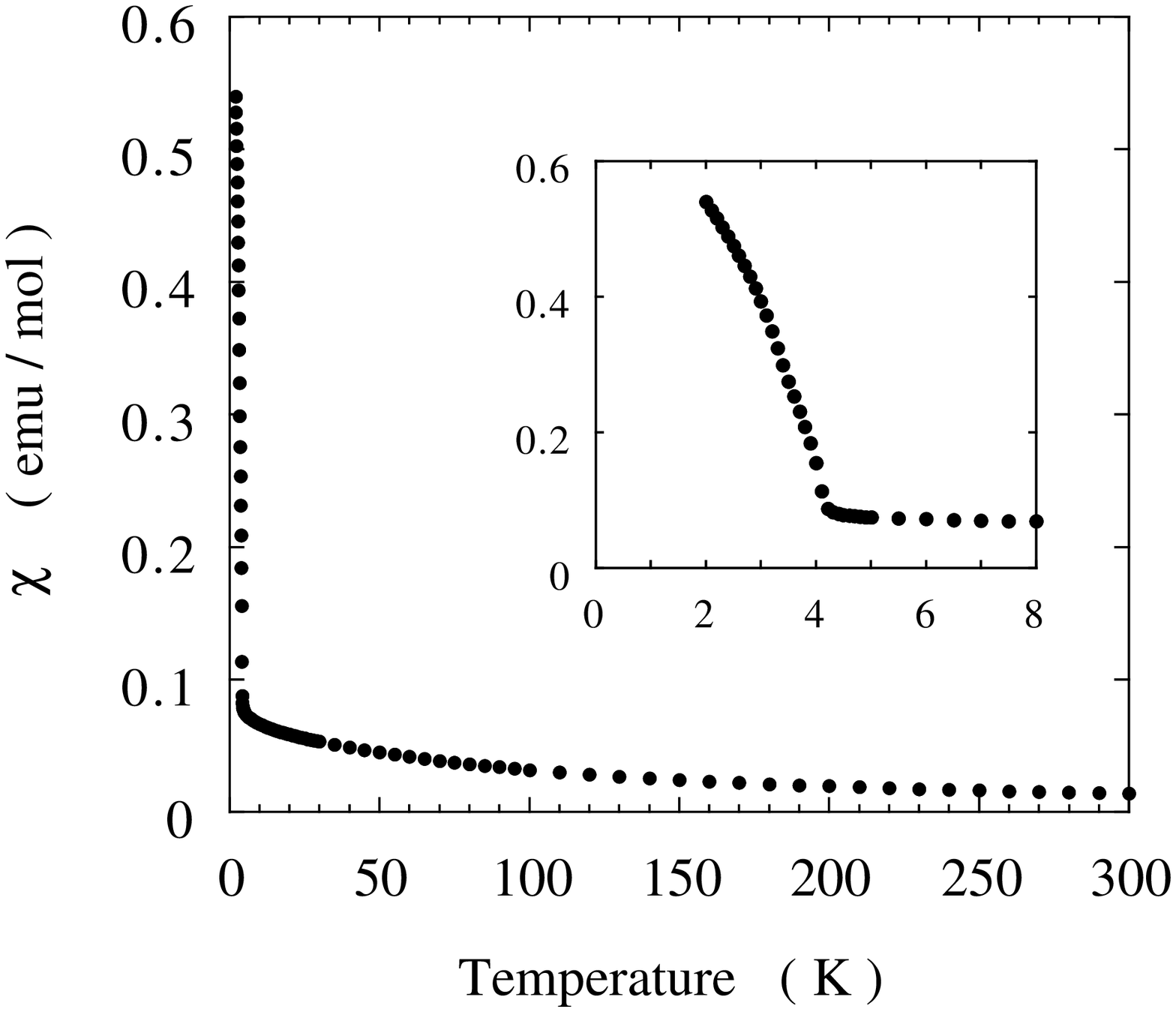,width=16cm}}
\vspace{0.5cm}
\caption{Temperature dependence of the susceptibility for
KCr$_3$(OH)$_6$(SO$_4$)$_2$.}
\label{MNFig2}
\end{figure}

The magnetization was measured using a SQUID magnetometer in the
temperature range between $2\,{\rm K}$ and $300\,{\rm K}$.
Figures \ref{MNFig1} and \ref{MNFig2} show the susceptibility of Fe
jarosite and Cr jarosite, respectively.
The susceptibility of Fe jarosite has the cusp at $T_{\rm{N[Fe]}} =
65\,{\rm K}$, while the susceptibility of Cr jarosite increases abruptly
below $4.2\,{\rm K}$.

\begin{figure}
\vspace{0cm}
\centerline{\epsfig{file=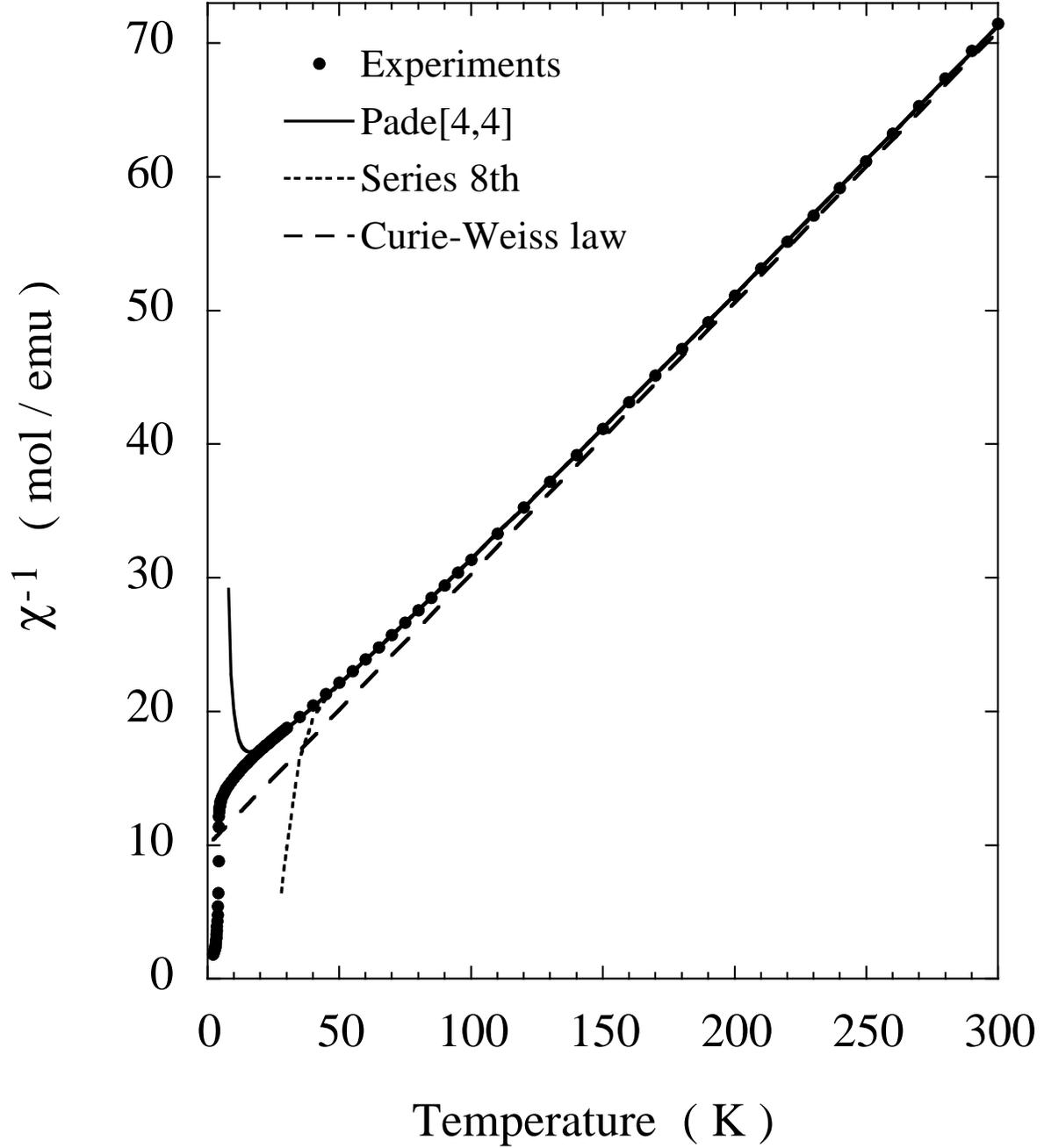,width=16cm}}
\vspace{0.5cm}
\caption{Observed and calculated inverse susceptibility of
KCr$_3$(OH)$_6$(SO$_4$)$_2$.
The closed circles are experimental data of $\chi^{-1}$ under the external
magnetic field $500\,{\rm Oe}$.
The result calculated from the high temperature series expansion up to the
8th order is shown by the dotted curve.
The result obtained by Pad\'{e} [4,4] approximants is shown by the solid curve.
The broken curve shows the Curie-Weiss law.}
\label{MNFig3}
\end{figure}

The susceptibility for the Heisenberg kagom\'{e} lattice antiferromagnet
has been calculated by the high temperature expansion up to the 8th order
and the result is extended to the lower temperature by the Pad\'{e} [4,4]
approximants \cite{RefHarris}.
Figure \ref{MNFig3} shows the comparison between the experimental and
theoretical inverse susceptibility for Cr jarosite.
The experimental values deviate clearly from the Curie-Weiss law below
about $150\,{\rm K}$ and fit very well with the calculated curves above
$20\,{\rm K}$.
The values of the exchange interaction for Cr jarosite and Fe jarosite are
derived to be $J_{\rm Cr}/k_{\rm B}= 4.9\,{\rm K}$ and $J_{\rm Fe} /k_{\rm
B}= 23\,{\rm K}$, respectively.
The theoretical susceptibility deviates remarkably from Curie-Weiss law
below about $8JS(S+1)/k_{\rm B}$ due to the development of short-range spin
correlation.
For Fe jarosite the value of $8JS(S+1)/k_{\rm B}$ is about $1600\,{\rm K}$,
that is far away from the experimental temperature region.
For the other kagom\'{e} lattice antiferromagnet
SrCr$_{8-x}$Ga$_{4+x}$O$_{19}$ the value is about $860\,{\rm K}$ and is
also large \cite{RefHarris,RefSCGO}.
The value for Cr jarosite is about $150\,{\rm K}$, that is in the
experimental temperature region.
Thus Cr jarosite is a good example to show the deviation from the
Curie-Weiss law owing to the small $J$ and $S$ values.

The $^1$H-NMR spectrum of Fe jarosite has a sharp peak in the paramagnetic
phase, while the spectrum below $65\,{\rm K}$ becomes to be broader and
shows typical pattern for the powder antiferromagnets \cite{RefAFMshape}.
This transition temperature coincides with the susceptibility data and the
spectrum indicates the antiferromagnetic ordering.

\begin{figure}
\vspace{0cm}
\centerline{\epsfig{file=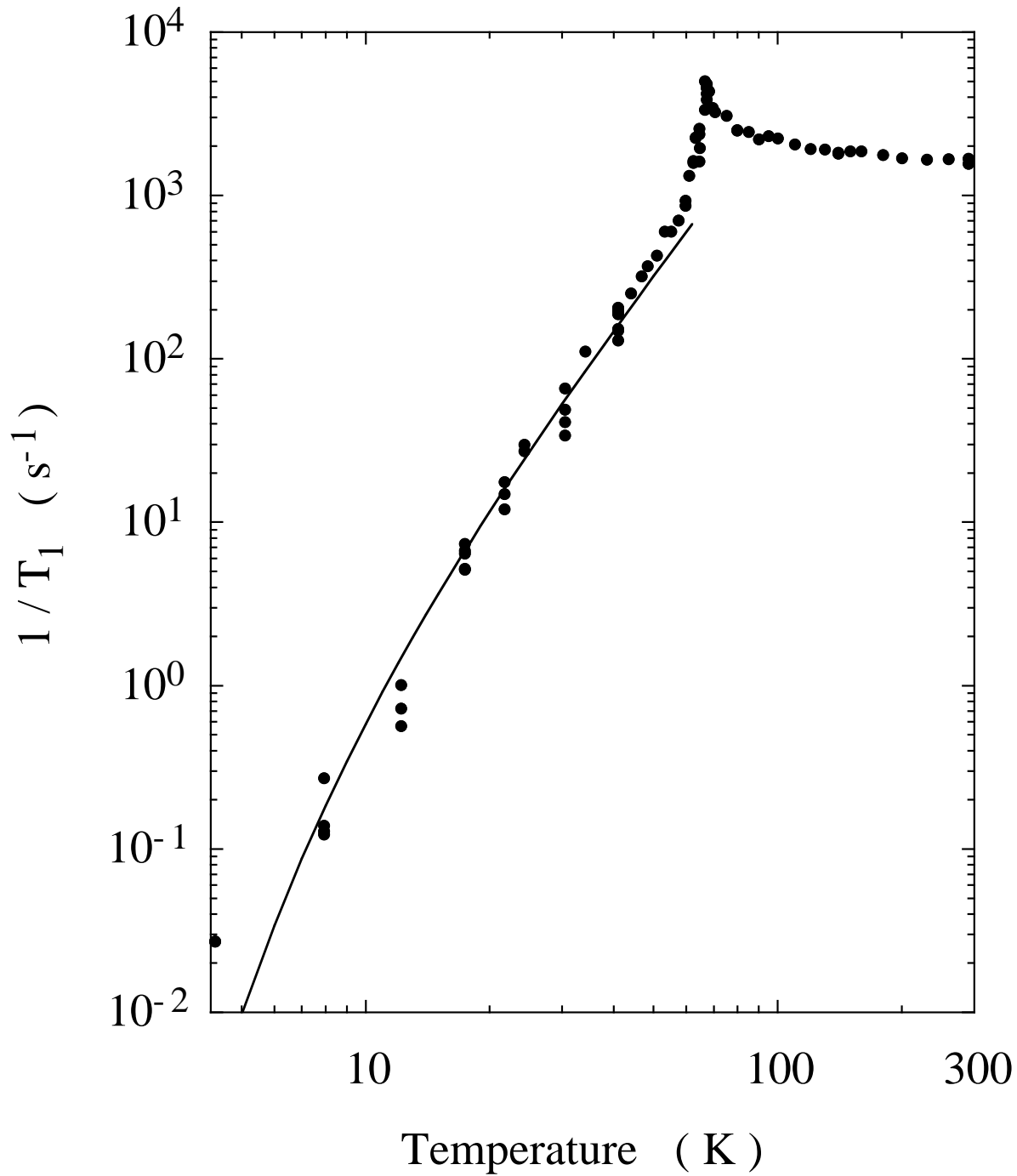,width=16cm}}
\vspace{0.5cm}
\caption{Proton spin-lattice relaxation rate of KFe$_3$(OH)$_6$(SO$_4$)$_2$.
The closed circles are experimental data at $75.1\,{\rm MHz}$.
The solid curve shows the temperature dependence of $1/T_1$ calculated from
the two-magnon process using the energy gap of $25\,{\rm K}$.}
\label{MNFig4}
\end{figure}

The spin-lattice relaxation rates, $1/T_1$, of $^1$H in Fe jarosite is
shown in Fig.~\ref{MNFig4}.
The rate $1/T_1$ in the paramagnetic phase slightly increases as
temperature approaches to $T_{\rm {N[Fe]}}$.
The rate in the ordered phase decreases sharply as temperature is lowered.

The NMR spectrum for Cr jarosite has a sharp peak in the paramagnetic
phase, while the half width increases below $4.2\,{\rm K}$.
The rate $1/T_1$ for Cr jarosite is almost independent of the temperature
in paramagnetic phase, however, it decreases below $4.2\,{\rm K}$ as
temperature is lowered.
\section{DISCUSSION}
The NMR spectrum of Fe jarosite indicates that all protons feel same
magnitude of internal dipolar field from Fe$^{3+}$ spins.
This suggests that the ordered spin structure is the ${\bf q}\!=\!0$ type
of the $120^{\circ}$ configuration with positive chirality.
If there existed the magnetic alignment with negative chirality, two kinds
of proton sites with different magnitude of the internal field must exist.
The neutron diffraction experiments confirmed this magnetic structure and
revealed that the spins direct to or from the center of triangle on the
kagom\'{e} lattice \cite{RefInami}.

This spin structure is considered to be caused by the single-ion type
anisotropy of magnetic ions.
Each magnetic ion is surrounded by an octahedron composed of six oxygens,
whose principal axis cants about $\theta = 20^{\circ}$ from the $c$-axis
towards the center of a triangle and the octahedron deforms slightly.
The deformation and the canting of the octahedron must cause the single-ion
anisotropy.
The spin system can be expressed as,
\begin{equation}
{\cal{H}}=2J\!\!\sum_{<i,j>}\!{\bf{S}}_i\!\cdot\!{\bf{S}}_j
+D\sum_{i}(S_i^{z'})^{2}
-E\sum_{i}\{(S_i^{x'})^2-(S_i^{y'})^2\},
\label{eq:Hamiltonian}
\end{equation}
where the local coordinate $(x', y', z')$ for each ion is determined by the
relation with each surrounding octahedron, $D > 0$ and $E > 0$.
In the case of
\begin{equation}
E > D \frac{\sin^2 \theta }{1+\cos^2 \theta },
\label{eqED}
\end{equation}
the spin structure with the minimum energy for the system is the ${\bf
q}\!=\!0$ type with the positive chirality and the spins direct to or from
the center of the triangle on the kagom\'{e} lattice.
This means that the system corresponds effectively to the two-dimensional
Ising antiferromagnet, which has the magnetic phase transition at finite
temperature \cite{RefInami}.
The ordering in the plane would induce the three-dimensional ordering due
to the infinitesimal interplane interaction.

The relaxation rate in the ordered phase can be analyzed by the two-magnon
process of the spin wave in the Heisenberg kagom\'{e} lattice
antiferromagnet.
The relaxation rate by the two-magnon process is expressed as \cite{RefMoriya},
\begin{equation}
\frac{1}{T_1}=\frac{\pi}{2}\gamma_e^2\gamma_n^2\hbar^2\sum_{i,j}G_{ij}
\int_{\omega_0}^{\omega_m} \Biggl\{ 1+\biggl( \frac{\omega_m}{\omega}
\biggl)^2
\Biggl\}\frac{e^{\hbar\omega/k_{\rm{B}}T}}{(e^{\hbar\omega/k_{\rm{B}}T}-1)^2
}N(\omega)^2 \rm{d}\omega,
\label{eq2mag}
\end{equation}
where $G_{ij}$ is the geometrical factor of the dipolar interaction,
$\omega _m$ is the maximum frequency, $\omega _0$ is the energy gap and
$N(\omega)$ is the state density of magnons.
The dispersion relation of magnons in the system of ${\bf q}\!=\!0$ type
spin structure has been obtained by Harris et al. \cite{RefHarris}.
We adapt their method for Fe jarosite by introducing the anisotropy.
The dispersion curves have the energy gaps due to the anisotropy and the
lowest energy gap is given as,
\begin{equation}
\Delta \epsilon = S \sqrt{\Bigl(3J+2E+2 \bigl(E\cos^2\theta-D\sin^2\theta
\bigr)\Bigr)\Bigl(2 \bigl(D-E \bigr)+4 \bigl(E\cos^2\theta-D\sin^2\theta
\bigr)\Bigr)}.
\label{eqDC}
\end{equation}
Applying the long wave approximation for the dispersion relation, the
relaxation rate is given as \cite{RefMaegawaT1},
\begin{equation}
\frac{1}{T_1}=\frac{\pi}{2}\gamma_e^2\gamma_n^2\hbar^2\sum_{i,j}G_{ij}
\frac{9\hbar}{k_{\rm{B}}}\frac{1}{(T_m^2-T_0^2)^3}T^5
\int_{T_0/T}^{T_m/T} \Biggl\{ x^2-\Biggl( \frac{T_0}{T} \Biggl)^2
\Biggl\}\Biggl\{ x^2+\Biggl( \frac{T_m}{T} \Biggr)^2
\Biggl\}\frac{e^{x}}{(e^{x}-1)^2}{\rm{d}}x,
\label{eq2magT5}
\end{equation}
where $T_m=\hbar\omega_m/k_{\rm B}$ and $T_0=\hbar\omega_0/k_{\rm B}$.
We calculated the temperature dependence of $1/T_1$ and the calculated
values are shown in Fig.~\ref{MNFig4} by the solid curve.
The agreement between the experimental data and the calculated values is
fairly well and we get the value of energy gap to be $25\,{\rm K}$.
The values of $E$ and $D$ are estimated by using Eq. (\ref{eqED}) and
(\ref{eqDC}) as,
\begin{eqnarray}
0.0012<\frac{E}{J}< 0.020,\\
0   <\frac{D}{J}<0.020.
\label{eq:EDvalue}
\end{eqnarray}

\begin{figure}
\vspace{0cm}
\centerline{\epsfig{file=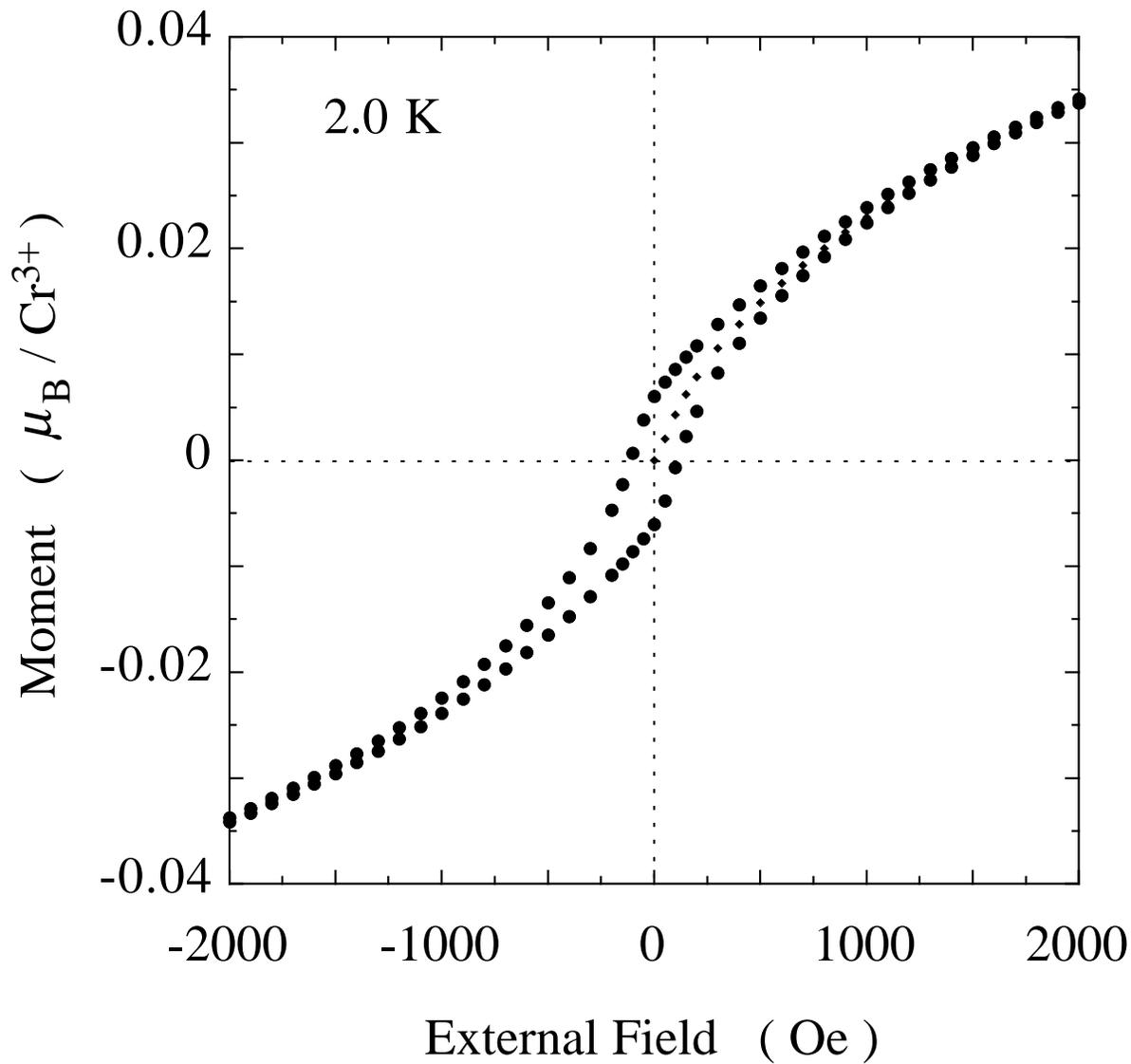,width=16cm}}
\vspace{0.5cm}
\caption{Magnetization curve of KCr$_3$(OH)$_6$(SO$_4$)$_2$ at $2.0\,{\rm K}$.}
\label{MNFig5}
\end{figure}

As is seen in Fig.~\ref{MNFig2}, the susceptibility for Cr jarosite
increases sharply below $4.2\,{\rm K}$.
Below this temperature the difference between the susceptibility measured
after the zero-field cooling (ZFC) and that measured after the field
cooling (FC) was observed.
The same behavior has been reported by  A. Keren {\it{et al}}. \cite{RefKeren}.
The magnetization curve was measured at 2.0 K to clarify this anomaly and
is shown in Fig.~\ref{MNFig5}.
There we find a small hysteresis loop, which suggests the existence of weak
ferromagnetic moments.
The difference in the susceptibility between ZFC and FC comes from the
hysteresis loop below $4.2\,{\rm K}$.
By our neutron diffraction experiment for Cr jarosite the long-range
ordering has been observed below $4.2\,{\rm K}$.
S. -H. Lie {\it{et al}}{.} have also reported the weak long range
antiferromagnetic ordering observed by the neutron experiments
\cite{RefSHLie}.
We conclude that the transition of Cr jarosite at $4.2\,{\rm K}$ is not a
spin glass like but magnetic one.

The weak ferromagnetic moment is considered to be caused by the canting of
the $120^{\circ}$ arrangement perpendicular to the $c$-plane due to the
anisotropy.
On the other hand, the ferromagnetic moment was not observed for Fe jarosite.
These results can be explained by the antiparallel stacking of the net
moments on the $c$-plane for Fe jarosite and parallel stacking of the net
moments for Cr jarosite.
This is consistent with the result from the neutron experiments that the
magnetic unit cell in Cr jarosite is equal to the chemical unit cell, while
that in Fe jarosite is double the chemical unit cell along the $c$-axis.
\section{SUMMARY}
In summary we have investigated the kagom\'{e} lattice antiferromagnets
KFe$_3$(OH)$_6$(SO$_4$)$_2$ and KCr$_3$(OH)$_6$(SO$_4$)$_2$ by means of the
magnetization, NMR and neutron experiments.
The susceptibility data of these samples are well fitted with the
susceptibility calculated by the high temperature expansion for the
two-dimensional Heisenberg kagom\'{e} lattice antiferromagnet.
Long range magnetic ordering occurs at 65 K for
KFe$_3$(OH)$_6$(SO$_4$)$_2$.	The spin structure in the ordered phase is
the $120^{\circ}$ structure with ${\bf{q}}\!=\!0$, +1 chirality and the
direction being to or from the center of the triangle.
The order is caused by the single-ion-type anisotropy $D, E$.
This system corresponds effectively to the two-dimensional Ising
system, which has the two-dimensional ordering.
The fluctuation in the ordered phase is caused by the spin wave.
The nuclear spin-lattice relaxation is governed by the two-magnon process.
For KCr$_3$(OH)$_6$(SO$_4$)$_2$ magnetic transition occurs at 4.2 K.
The transition is not spin-glass one but the magnetic ordering.
The weak ferromagnetic moments are observed.
This comes from the canted $120^{\circ}$ spin structure,
and all net moments in the planes are parallel to the $c$-axis.
\end{document}